\documentclass[journal,10pt]{IEEEtran}
\makeatletter
\def\ps@headings{%
\def\@oddhead{\mbox{}\scriptsize\rightmark \hfil \thepage}%
\def\@ adversarynhead{\scriptsize\thepage \hfil \leftmark\mbox{}}%
\def\@oddfoot{}%
\def\@ adversarynfoot{}}
\makeatother
\ifCLASSINFOpdf
\else

\fi

\newcounter{problem}
\newcounter{save@equation}
\newcounter{save@problem}

\makeatletter

\usepackage{epsfig}
\usepackage{array}
\newcolumntype{L}[1]{>{\raggedright\let\newline\\\arraybackslash\hspace{0pt}}m{1}}
\newcolumntype{C}[1]{>{\centering\let\newline\\\arraybackslash\hspace{0pt}}m{1}}
\newcolumntype{R}[1]{>{\raggedleft\let\newline\\\arraybackslash\hspace{0pt}}m{1}}
\usepackage[short]{optidef}
\usepackage{amsthm} 
\usepackage{mathrsfs}
\newcommand{\bc}{\begin{center}}
\newcommand{\ec}{\end{center}}
\newcommand{\be}{\begin{equation}}
\newcommand{\ee}{\end{equation}}
\usepackage{flexisym}
\usepackage{algorithmicx}
\usepackage{algorithm}
\usepackage{siunitx}
\usepackage{comment}
\usepackage{soul}
\usepackage{algpseudocode}

\newcommand{\bnu}{\begin{enumerate}}
\newcommand{\enu}{\end{enumerate}}
\usepackage{cite}
\usepackage{url}
\usepackage[utf8]{inputenc}
\usepackage[T1]{fontenc}
\usepackage{amsmath}
\usepackage{amsmath}

\usepackage{amsfonts}
\usepackage{amssymb}
\usepackage[version=4]{mhchem}
\usepackage{stmaryrd}
\usepackage{graphicx}
\usepackage[export]{adjustbox}
\graphicspath{ {./images/} }
\usepackage{breqn}
\usepackage{lipsum}
\usepackage{mathtools}
\usepackage{caption}
\usepackage{float}
\usepackage{subcaption}
\usepackage{soul}
\usepackage{blindtext, graphicx}
\usepackage{cuted}
\usepackage{color}

\usepackage{amsfonts}
\usepackage{multicol}
\usepackage{lipsum}
\usepackage{cuted}

\usepackage{graphicx}
\graphicspath{ {./images/} }

\newtheoremstyle{case}{}{}{}{}{}{:}{ }{}

\begin{document}
\title{Adaptive Context-Aware Multi-Path Transmission Control for VR/AR Content: A Deep Reinforcement Learning Approach}
\author{Shakil Ahmed,~\IEEEmembership{Member,~IEEE}, Saifur Rahman Sabuj,~\IEEEmembership{Member,~IEEE}, and Ashfaq Khokhar,~\IEEEmembership{Fellow,~IEEE}
\vspace*{-0.80 cm}
\thanks{Shakil Ahmed and Ashfaq Khokhar are with the Department of Electrical and Computer Engineering, Iowa State University, Ames, Iowa, USA. (email: \{shakil, ashfaq\}@iastate.edu). S. Sabuj is with the Department of Electrical and Electronic Engineering, BRAC University, Bangladesh. (email: srsabuj@bracu.ac.bd) }
}

{}

\maketitle

\begin{abstract}
This paper introduces the Adaptive Context-Aware Multi-Path Transmission Control (ACMPTC) system, an efficient approach designed to optimize the performance of Multi-Path Transmission Control Protocol (MPTCP) for data-intensive applications such as augmented and virtual reality (AR/VR) streaming. ACMPTC addresses the limitations of conventional MPTCP by leveraging deep reinforcement learning (DRL) for agile end-to-end path management and optimal bandwidth allocation, facilitating path realignment across diverse network environments. The primary contributions of this system include comprehensive bandwidth requirement analysis, feedback mechanisms, congestion forecasting, and adaptive data routing, all of which contribute to low latency and high network utilization. A mathematical model has been developed to validate the DRL-based ACMPTC framework's effectiveness, which optimizes path selection, bandwidth allocation, and congestion control under varying and unpredictable network conditions. Extensive simulations demonstrate that ACMPTC outperforms competing MPTCP schemes in terms of throughput, latency, and quality of service, making it a promising candidate for new network connection management in AR/VR applications and adaptive network traffic management techniques.
\end{abstract}

\vspace*{-0.25 cm}
\section{Introduction} 
The rapid evolution of data-intensive applications, particularly in augmented and virtual reality (AR/VR), has introduced unprecedented demands on network infrastructures \cite{cisco2020annual}. These applications require high bandwidth, ultra-low latency, and consistent quality of service (QoS) to deliver seamless, immersive experiences \cite{elbamby2018toward}. Traditional network protocols like the Transmission Control Protocol (TCP) often struggle to meet these stringent demands, especially in highly dynamic and diverse network environments due to single path transmission, inadequate for high-bandwidth, low-latency requirement, high latency sensitivity, etc. \cite{li2016multipath}. These limitations make TCP less effective for dynamic, heterogeneous network environments and the demanding performance needs of modern applications like AR/VR, which require more adaptive and efficient data transmission mechanisms.

A potential solution that has gained in popularity to cope with the demands of high-bandwidth, low-latency applications is Multi-Path TCP (MPTCP), which allows multiple network paths between two hosts being used simultaneously \cite{ford2013tcp}. Although MPTCP has improved performance to some extent in terms of bandwidth, it is slow at responding to dynamic network conditions (e.g., may perform inefficiently and unbalance the links) with a lack of context awareness as well are less capable of congestion control, which makes them an unsuitable solution for increased-bandwidth applications i.e., AR/VR etc. These shortcomings restrict the ability of MPTCP to satisfy stringent performance needs for AR/VR — rapidly adapting to network changes and efficiently managing bandwidth are essential for a seamless, high-quality user experience.
To address the limitations of TCP and MPTCP, this paper proposes the Adaptive Context-Aware Multi-Path Transmission Control (ACMPTC) system for AR/VR content. ACMPTC advances current MPTCP solutions by integrating deep reinforcement learning (DRL) techniques for intelligent and dynamic path management and bandwidth allocation \cite{mao2016resource}. This novel approach enables ACMPTC to respond rapidly to varying network condtions, critical to maintaining high-quality AR/VR streaming experiences.

The difference is that ACMPTC can adjust data transmission paths based on network conditions. ACMPTC employs a feedback mechanism, path-selection latency minimization, and network-utilizing maximizable optimization \cite{xu2018experience}. Per the DRL-based solution, a multi-agent with higher autonomy levels continuously monitors and controls network paths on its behalf based on real-time decisions, like path selection, congestion control roles, as well bandwidth limiting. In this article, the authors present a few critical features for ACMPTC to enhance its performance—mainly choosing paths with low latency and packet loss. It brings a DRL-based agent that can adapt its decision to real-time network states and compute dynamic, optimal choices. This feedback loop, on the other hand, allows for real-time path selection and resource allocation that enables continuous optimization to provide a smooth AR/VR experience even with varying network conditions. It confirms that the system operates correctly and provides a way to update such a network when there is variation in traffic levels by adjusting it effectively.

\subsection{Background and Significance} The rise of multi-path networking technologies and high-bandwidth hungry AR/VR brings complexity to managing network traffic. However, it also offers opportunities to improve network performance and ensure uninterrupted immersive experiences. To address these challenges, we propose ACMPTC, which dynamically adapts based on network conditions and optimizes path selection for AR/VR applications.
In AR/VR multimedia applications, QoS is sensitive to latency, bandwidth availability, and packet loss. ACMPTC plays a crucial role by minimizing latency and ensuring efficient path selection, improving network performance and the overall QoS \cite{torres2020immersive}.

The most significant benefit of ACMPTC is its capability to learn on the fly and optimize bandwidth across what they say are more appropriate paths. However, network congestion management and maintaining consistent communication has become more complex, especially in urban areas, as the traffic volume has increased \cite{rojviboonchai2004evaluation}. ACMPTC tackles those challenges using a real-time path management mechanism that adjusts to these dynamic network conditions without causing service disruption. This system is essential for maintaining service during peak times.
Factors like ACMPTC's dynamic adaptability and context awareness make it efficient to present a stable content delivery in an AR/VR environment \cite{amir1995efficient}. It also looks for possible QoS degradations and redistributes the available resources to ensure QoS for the resource-hungry R/VR content.
Additionally, ACMPTC boasts its DRL framework that acts in a full-automatic manner to handle congestion and optimize bandwidth allocation. DRL-based prediction of congestion and periodic proactive resource (bandwidth) allocation using up-to-date network data by the DRL algorithm guarantees that bandwidth is not taken away from AR / VR applications without delay [8683970]. This innovative, principled structure of ACMPTC makes it a potential technology for 6G demand-driven network traffic management, guaranteeing user experience with passivity.

\subsection{Related Work and Contribution} \label{RW}
Understanding the evolution of network protocols and optimization strategies is essential for advancing multi-path data transmission technologies. Our literature review spans the development of MPTCP and the integration of DRL for network optimization, identifying areas where further research can contribute to the field.

\subsubsection{Evolution of MPTCP}
The inception of MPTCP marked a paradigm shift in network protocol design aimed at leveraging the increasing availability of diverse network paths to enhance data transmission reliability and throughput. Initial theoretical explorations in \cite{liu2015improving} provided a compelling case for MPTCP, indicating potential benefits such as improved bandwidth utilization and failover capabilities. Practical implementations and subsequent challenges, particularly in path management and congestion control, were later discussed \cite{frommgen2016remp}, highlighting the intricacies of applying MPTCP in real-world networking scenarios.
Over time, research has demonstrated MPTCP's adaptability to different network architectures, from wired broadband connections to mobile networks \cite{xu2002tcp}. Moreover, the adoption of MPTCP in mainstream operating systems and its implications for application performance have been rigorously analyzed \cite{chuachan2018solving}, revealing its potential and limitations in handling today's network traffic demands.

\subsubsection{Advances in DRL for Network Optimization} 
The inception of MPTCP marked a paradigm shift in network protocol design aimed at leveraging the increasing availability of diverse network paths to enhance data transmission reliability and throughput. Primarily, network decisions depended on static routing and management policies \cite{xu2018experience}. However, due to the nature of network states, which change rapidly, it is needed to have more flexible and adaptive solutions. This is where DRL started revolutionizing network management by allowing networks to learn and improve with more experience. The authors in \cite{xu2018experience} are among the first to adopt DRL for network decisions. They used it to change the routing paths on the fly, enabling them to make a significant breakthrough in dealing with changing network topologies and congestion patterns. They unlocked the door from static setups, allowing DRL to take proactive, context-aware decisions leading to optimal data flow \cite{zhang2001tcp}.

 DRL efficiently handles high-dimensional state spaces and complex decision-making environments. DRL is ideal for handling vast amounts of network data and generating policies best suited to real-time network conditions — hence, it grounds its TCP as a type of DRL. However, the full potential of DRL in the context of network operation goes beyond the routing itself \cite{wang2023integrating}. Optimizing multi-path routing with DRL is possible across various network operation aspects, such as load balancing, resource allocation, or quality of service optimization. The aspects of DRL implementation in network systems are pretty challenging. Namely, network conditions are dynamic and are not always statistically the same. Due to high variability, DRL systems must converge quickly and provide a stable solution. Furthermore, network conditions may not always be stationary; therefore, a separate solution is needed to allow DRL systems to adapt to novel situations without human intervention \cite{silva2018mptcp}.

\subsubsection{Gaps in Current Research}
Prior works provided foundational studies using MPTCP and DRL in network systems. However, optimization of the two for AR and MPTCP technologies, which are bandwidth and ultra-low latency demanding, remains an area of MPTCP’s applicability but needs significant exploration. The existing network optimization, mostly reactive, based on DRL general algorithms, cannot anticipate the best network conditions prior, consequently heavily impacting AR/VR streaming quality \cite{wang2018proactive}. The capacity to predict future user states in DRL encourages their use in the current research as a promising approach to advance MPTCP usage \cite{wang2023integrating}. With this unified framework lacking in the above studies, the gap between theoretical advancements and their implementation in viable solutions overshadow works on this phenomenon in the current study \cite{silva2018mptcp, wang2018proactive, wang2023integrating}. Our study attempts to bridge this theoretical-practical gap by presenting a complete system that optimizes ACMTCP behavior while enhancing the efficiency of DRL in achieving AR/VR demands with high bandwidth and low latency. This work further bridges the gap in network optimization theory. It also validates the resulting solutions’ practicality in real-world networks to advance further network protocol design capabilities aligned with evolving digital application trends. 

\subsection{Contributions}
Our work is guided by the following critical achievements aimed at enhancing MPTCP for current data-intensive applications such as AR/VR streaming:

\begin{itemize}
  \item The primary contribution is developing the ACMPTC system, which dynamically optimizes multi-path networking by adjusting path selection and bandwidth allocation across available channels. Unlike traditional approaches, ACMPTC leverages real-time network metrics to forecast bandwidth demands and allocates/reallocates the most efficient paths to ensure seamless AR/VR streaming.

\item We introduce a state-of-the-art DRL-based framework within ACMPTC, where a multi-agent monitors and controls the network paths. The agent autonomously makes optimal decisions on path selection, congestion control, and bandwidth distribution, enabling quick adaptation to network fluctuations. This is especially critical for maintaining quality in high-speed, low-latency AR/VR applications.

\item In the ACMPTC system, we design comprehensive state, action, and reward functions that empower the DRL agents to learn and execute effective strategies for network management. The ACMPTC model ensures robust and efficient AR/VR streaming by systematically optimizing these network control parameters. A mathematical model has also been developed to validate the system's operations, supporting adaptive responses to varying network conditions crucial for high-bandwidth and low-latency multimedia applications.

\item A significant enhancement in ACMPTC is incorporating a feedback mechanism, enabling continuous optimization based on real-time network performance metrics. This feedback loop, driven by the insights and adjustments from the DRL agents, dynamically reallocates paths and resources for AR/VR content, effectively responding to changes in network conditions. This adaptive architecture ensures a consistent and high-quality QoS by minimizing disruptions.

\item The proposed ACMPTC system provides a comprehensive evaluation framework through simulation, outperforming traditional TCP and MPTCP management techniques. By maintaining steady data transmission quality, ACMPTC significantly improves QoS in AR/VR streaming environments, ensuring reliable and smooth content delivery. 
\end{itemize}

\textit{Paper Scope and Structure:}
Section~\ref{TF} defined the fundamentals of the MPTCP model. Section~\ref{AMF} details the mathematical constructs of ACMPTC model. This paper is further structured as follows: Section~\ref{Section:ProposedSolution} illustrates the proposed DRL-based solution. Section~\ref{SE} presents the simulation procedures and the evaluations, demonstrating how ACMPTC performs in controlled and realistic situations. Section~\ref{C} presents the research findings and discusses the implications and potential for future network communications of the ACMPTC system.

\section{Theoretical Framework: MPTCP} \label{TF}
\begin{figure}[]
\centering
\includegraphics[width=3.2in]{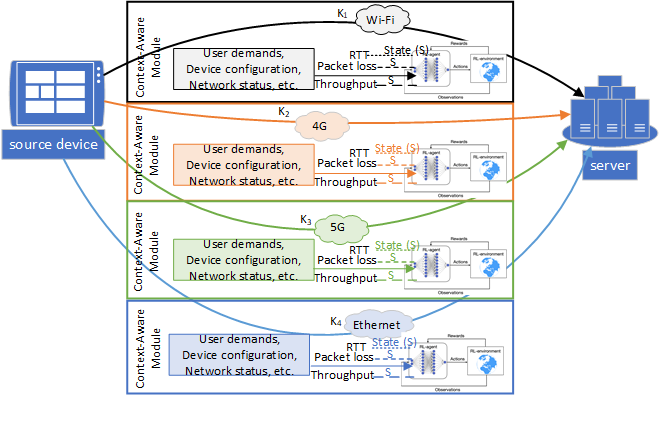}
\caption{The source device utilizes multiple network paths (Wi-Fi, 4G, 5G, Ethernet) to send data to the server. The MPTCP protocol optimizes the use of these paths by managing traffic across sub-flows, improving throughput and fault tolerance.}
\label{Fig_MPTCP}
\end{figure}

We provide the theoretical framework for MPTCP that extends the traditional TCP protocol to utilize multiple network paths simultaneously, increasing overall throughput and improving resilience against network failures \cite{mptcp2013}. Using multiple sub-flows across different paths allows MPTCP to distribute data more efficiently across network resources.

\subsection{ MPTCP sub-flows and Throughput}
An MPTCP connection is comprised of multiple sub-flows, denoted as ${K_1, K_2, \ldots, K_n}$, where \( n \) represents the total number of sub-flows, each corresponding to a different network path. Each sub-flow \( K_i \) operates as a single TCP connection over one of the available paths (e.g., Wi-Fi, 4G, 5G, Ethernet, etc.), as shown in Fig.~\ref{Fig_MPTCP}.
The total throughput of the MPTCP connection, \( T_{\text{total}} \), is the sum of the throughputs of individual sub-flows:
\begin{equation}
T_{\text{total}} = \sum_{i=1}^{n} T_{K_i}
\end{equation}

Following the model proposed in \cite{emptcp2024}, the throughput of each sub-flow \( K_i \) can be modeled as:
\begin{equation}
\label{Eq_subf}
T_{K_i} = \frac{w_i}{RTT_i} \left(1 - \frac{p_i}{2}\right)
\end{equation}
where  \( w_i \) is the window size of sub-flow \( K_i \),
 \( RTT_i \) is the round-trip time of sub-flow \( K_i \), and
  \( p_i \) is the packet loss probability on sub-flow \( K_i \).
Note (\ref{Eq_subf}) captures how sub-flow the window size, latency, and packet loss on each path influence performance.

\subsection{Traffic Allocation and Optimization}
The optimization of MPTCP focuses on finding the optimal allocation of traffic across the sub-flows to maximize overall throughput, subject to network constraints. The objective function is to maximize the total throughput \( T_{\text{total}} \):

\begin{subequations}
\begin{align}
  \max_{\{D_{K_1}, \ldots, D_{K_n}\}} \quad & T_{\text{total}} = \sum_{i=1}^{n} \frac{w_i}{RTT_i}\left(1 - \frac{p_i}{2}\right) \\
  \text{s.t.} \quad & \sum_{i=1}^{n} D_{K_i} \leq D, \quad D_{K_i} \geq 0 \quad \forall i
\end{align}
\end{subequations}
where \( D \) is the total data load to be distributed across sub-flows, and \( D_{K_i} \) is the portion of data allocated to sub-flow \( K_i \).
The window size \( w_i \) can be modeled in equilibrium as:
\begin{equation}
\label{Eq_w_i}
w_i = \sqrt{\frac{2(1 - p_i)}{p_i}} \approx \sqrt{\frac{2}{p_i}}
\end{equation}
Eq.~(\ref{Eq_w_i}) shows how the window size adapts based on the packet loss probability across the sub-flows.
While MPTCP offers significant performance improvements, one of its key challenges is handling path failures or congestion on specific sub-flows, which can lead to performance degradation. MPTCP employs congestion control mechanisms to adjust window sizes and data distribution in response to changes in network conditions.
For instance, feedback-based path failure detection mechanisms have been proposed to rapidly detect path issues and adjust traffic routing accordingly \cite{feedback2016}. These mechanisms help mitigate packet loss and reallocate traffic to more stable sub-flows, improving the system’s overall resilience.

\subsection{Limitations of MPTCP}
Although MPTCP enhances throughput and fault tolerance by utilizing multiple paths, it faces limitations in handling highly dynamic network environments where real-time adaptation to network conditions is required. MPTCP’s standard congestion control algorithms may not be sufficient for complex applications like AR/VR streaming, where low latency and high bandwidth are critical. This sets the stage for further enhancement, which we will address in the following section by developing the ACMPTC model.

\section{ACMPTC Model} \label{AMF}
\begin{figure}[]
\centering
\includegraphics[width=3.2in]{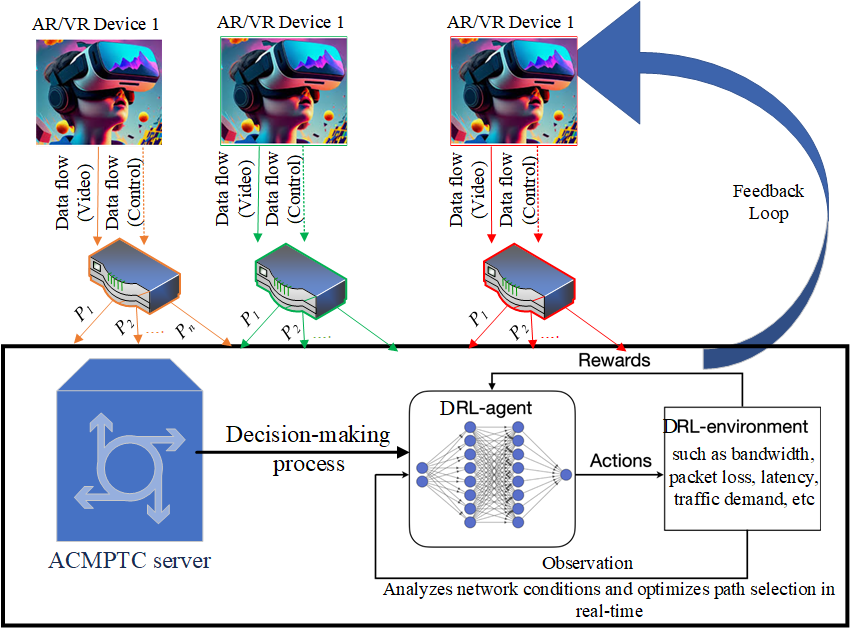}
\caption{AR/VR devices send data to the ACMPTC server,
which, with insights from the DRL agent, dynamically optimizes
data transmission paths to enhance AR/VR streaming quality}
\label{Fig_SystemModel}
\end{figure}

To support a multi-path networking environment for 6G's advanced transmission needs, we introduce the ACMPTC system to optimize the performance of AR/VR multimedia applications, as depicted in Fig.~\ref{Fig_SystemModel}. ACMPTC system dynamically optimizes path selection, congestion control, bandwidth allocation, feedback mechanism, and context awareness related to AR/VR multimedia applications. Our model operates under the following assumptions:

\textit{1) Network State Dynamics:} We assume that the network state is inherently dynamic, influenced by external conditions and fluctuations in user interaction. These fluctuations affect the state of each path, with a particular focus on AR/VR content delivery parameters.

\textit{2) Real-time Information and Feedback Loop:} AR/VR content is assumed to evolve dynamically. Consequently, ACMPTC leverages real-time insights and feedback to refine path selection and bandwidth allocation strategies continuously, optimizing network performance.

\textit{3) Continuous Adaptation and Learning:} The ACMPTC system incorporates a robust framework enabling continuous adaptation and learning. This ensures it evolves alongside emerging AR/VR technologies, optimizing network performance and maintaining the QoS.

\subsection{Network Model}
 ACMPTC framework is designed to support multi-path data transmission for delivering seamless, high-quality AR/VR streaming in dynamic and resource-constrained environments across various network paths, each with varying bandwidth, latency, congestion, and packet loss characteristics.
In the ACMPTC system, the available paths are denoted as \( \mathcal{P} = \{ P_1, P_2, \dots, P_n \} \), where each path \( P_i \) is characterized by dynamic metrics such as:
available bandwidth, \( B_i[t] \) on path \( P_i \) at time instant \( t \),
latency, \( L_i[t] \) on path \( P_i \), and 
congestion level, \( C_i[t] \) on path \( P_i \) at time \( t \).
Round-Trip Time \( RTT_i[t] \) measures the time it takes for a data packet to travel from the source to the destination and return with an acknowledgment (ACK).
For each AR/VR stream, \( S_j \), where  \( \{S_1, S_2, \ldots, S_j, \ldots, S_n\} \), transmitted over \( P_i \), it maintains a congestion window, noted as \( cwnd_{i,j}[t] \). The congestion window is that who dictates the number of active packets without being dropped at any given time. A larger congestion window enables faster data transfer. Even though the RTT defines an upper bound on how fast data can be acknowledged and therefore fixed kind of limits to the minimum send rate for the subsequent transmission.
In the following sub-sections, we model path selection, congestion control of AR/VR data stream, bandwidth allocation, feedback mechanism, utility function, and optimization problem for the ACMPTC model.

\subsection{Dynamic Path Selection, Assignment, and Adjustments}
In a multi-path network, the ACMPTC model also needs an efficient dynamic path selection opportunistic data transfer for AR/VR streaming. This subsection covers the path assignment. First, path selection approaches afterward, and then dealing with real-time network troubles such as congestion, RTT, bandwidth availability, and finally about, multi-path data transmission.

\subsubsection{Initial Path Assignment}
The system assigns each of the AR/VR stream \( S_j \) to one or more network paths \( \mathcal{P}_{S_j} \subseteq P\},\). The initial path assignment is based on an estimation of the current network conditions, such as \( B_i[t] \), \( L_i[t] \), \( RTT_i[t] \), and \( C_i[t] \). The system maps the AR/VR stream \( S_j \) to routes offering enough bandwidth and low latency, as real-time AR/VR content requires high performance. This process ensures the selected paths provide a reliable QoS for an immersive AR/VR experience.

\subsubsection{Dynamic Path Selection}
The ACMPTC system employs a dynamic path selection process to assign paths to the data streams based on real-time network conditions for optimal performance. AR/VR stream \( S_j \) is assigned a subset of paths \( \mathcal{P}_{S_j} \subseteq \mathcal{P} \), based on its bandwidth, RTT, and latency requirements as follows:
\begin{equation}
\label{Eq_path}
\mathcal{P}_{S_j}^*[t] = \arg\max_{\mathcal{P}_{S_j} \subseteq \mathcal{P}} \left\{ \sum_{P_i \in \mathcal{P}_{S_j}} \left( \alpha B_i[t] - \beta L_i[t] - \gamma \rho_i[t] \right) \right\},
\end{equation}
where $\mathcal{P}_{S_j}^*[t]$ defines the optimal set of paths selected for $S_j$ at $t$. \( \alpha, \beta, \gamma \) are weighting factors that prioritize bandwidth, RTT, latency, and packet loss, respectively. Note (\ref{Eq_path}) captures the paths selected for AR/VR stream, $S_j$ meet the AR/VR's quality requirements, with higher-weighted factors reflecting the priority of various parameters, such as minimizing latency and RTT for interactive AR/VR applications \cite{qosaware2013}.

\subsubsection{Dynamic Path Reallocation}
After an AR/VR $S_j $ stream is initiated, the ACMPTC model checks in on each available path to track real-time conditions and guarantee that it stays performant. As the network changes, such as high to low traffic, congestions, or increased RTT, paths may be loaded over their means to carry out and cause service degradation. To manage these challenges, the ACMPTC uses a dynamic path reallocation system. If the system detects a path is no longer suitable (e.g., due to congestion or increased latency), it dynamically reassigns the stream to alternative paths based on the following key factors:

\textit{Congestion Level:} The system continuously monitors the congestion level \( C_i[t] \) on path \( P_i \), which is the current path assigned to stream \( S_j \). If the congestion level exceeds a predefined threshold \( C_{\text{th}} \), i.e., \( C_i[t] > C_{\text{th}} \), the ACMPTC system reallocates traffic to another path with lower congestion. The optimal path selection for stream \( S_j \) is determined by the following:
\begin{equation}
\mathcal{P}_{S_j}^*[t] \!=\! 
\begin{cases}
\arg\min_{P_k \in \mathcal{P} \setminus \{P_i\}} \left\{ C_k[t] \right\},\!\!\!\!\! & \text{if} \ C_i[t] > C_{\text{th}} \\
P_i, & \text{if} \ C_i[t] \leq C_{\text{th}}
\end{cases}
\end{equation}
where \( \mathcal{P}_{S_j}^*[t] \) is the set of optimal paths selected for stream \( S_j \) at time \( t \).
\( P_i \) is the current path assigned to stream \( S_j \).
\( C_i[t] \) is the congestion level on the current path \( P_i \) at time \( t \).
\( C_{\text{th}} \) is the predefined congestion threshold.
\( \mathcal{P} \) is the set of all available paths.
 \( C_k[t] \) is the congestion level on an alternative path \( P_k \) at time \( t \).
The path reassignment process operates as follows:
If the congestion level on the current path \( P_i \) exceeds the threshold \( C_i[t] > C_{\text{th}} \), the system will reassign stream \( S_j \) to the path \( P_k \) with the lowest congestion level from the available paths.
 If the congestion level on the current path \( P_i \) is below or equal to the threshold \( C_i[t] \leq C_{\text{th}} \), \( S_j \) continues to use the current path \( P_i \) without reassignment.

\textit{RTT and Latency:} When \( RTT_i[t] \) or \( L_i[t] \) on \( P_i \) exceeds the thresholds \( RTT_{\text{th}} \) or \( L_{\text{th}} \), the ACMPTC reallocates AR/VR stream ${S_j}$ to other paths with lower RTT or latency:
\begin{equation*}
  \mathcal{P}_{S_j}^*[t] = 
  \begin{cases} 
    \arg\min_{P_k \in \mathcal{P} \setminus \{P_i\}} \left\{ RTT_k[t], L_k[t] \right\}, \\
    \quad \text{if} \ RTT_i[t] > RTT_{\text{th}} \ \text{or} \ L_i[t] > L_{\text{th}} \\
    P_i, 
    \quad \text{if} \ RTT_i[t] \leq RTT_{\text{th}} \ \text{and} \ L_i[t] \leq L_{\text{th}}.
  \end{cases}
\end{equation*}
where 
\( RTT_i[t] \) is the round-trip time on the current path \( P_i \) at time \( t \).
\( L_i[t] \) is the latency on the current path \( P_i \) at time \( t \).
 \( RTT_{\text{th}} \) and \( L_{\text{th}} \) are the predefined RTT and latency thresholds, respectively.
 \( \mathcal{P} \) is the set of all available paths.
 \( RTT_k[t] \) and \( L_k[t] \) are the RTT and latency on an alternative path \( P_k \) at time \( t \).
 If the RTT or latency on the current path \( P_i \) exceeds the threshold \( (RTT_{\text{th}} \text{ or } L_{\text{th}}) \), the system reallocates \( S_j \) to the path with the minimum RTT or latency among the other available paths.
If the RTT and latency on \( P_i \) are below or equal to the thresholds, \( S_j \) will continue using the current path \( P_i \).

\textit{Packet Loss:} The ACMPTC system continuously monitors the packet loss rate on the current path \( P_i \) assigned to stream \( S_j \). If the packet loss rate \( \rho_i[t] \) on \( P_i \) exceeds the predefined threshold \( \rho_{\text{th}} \), the system reallocates the traffic from \( S_j \) to a different path with a lower packet loss rate. The path assignment is determined as:
\begin{equation}
  \mathcal{P}_{S_j}^*[t] = 
  \begin{cases}
    \arg\min_{P_k \in \mathcal{P} \setminus \{P_i\}} \left\{ \rho_k[t] \right\}, & \text{if} \ \rho_i[t] > \rho_{\text{th}} \\
    P_i, & \text{if} \ \rho_i[t] \leq \rho_{\text{th}}
  \end{cases}
\end{equation}
where 
\( \rho_i[t] \) is the packet loss rate on the current path \( P_i \) at time \( t \).
\( \rho_{\text{th}} \) is the predefined packet loss threshold.
\( \rho_k[t] \) is the packet loss rate on an alternative path \( P_k \) at time \( t \).
If the packet loss rate on the current path \( P_i \) exceeds the threshold \( (\rho_{\text{th}}) \), the system reallocates \( S_j \) to the path with the lowest packet loss rate among the other available paths.
- If the packet loss rate on \( P_i \) is below or equal to the threshold, \( S_j \) will continue using the current path \( P_i \).

\subsubsection{Handling Insufficient Bandwidth and Path Sharing}
For instance, an AR/VR stream \(S_j\) may have bandwidth requirements that cannot be supported entirely on its assignment paths. The ACMPTC system will automatically reshape the path assignments by adding more paths to support higher demand. On the other hand, if bandwidth requirements decrease (such as when AR/VR content complexity drops temporarily), the system will remove some assigned paths to normalize network resource consumption. When network resources are limited, and multiple AR/VR steams\( \{S_1, S_2, \ldots, S_j, \ldots, S_n\} \) need to share the same path, a common technique is known as Path Sharing \cite{6805833}. To that end, the system monitors path congestion level \(C_i[t]\) and adaptively allocates bandwidth among streams to ensure each stream performs satisfactorily.

\subsubsection{Multi-Path Data Transmission}
Multi-path transmission allows each AR/VR data stream \( S_j \) to be transmitted across one or more available paths \( P_i \). The system can balance the load, reduce latency, and mitigate packet loss by distributing the data across multiple paths. Paths are selected based on real-time performance metrics, including bandwidth, RTT, congestion window, and packet loss. This multi-path approach ensures robust traffic distribution, allowing the system to maintain continuous service even when some paths experience degradation or failure, a critical capability in dynamic network environments \cite{feedback2016}.

\subsection{AR/VR Data Stream-Aware Congestion Control}
The ACMPTC system includes an efficient congestion mechanism for AR/VR data streams, aiming to deliver low latency, high throughput, and minimal packet loss to ensure a smooth QoS. This mechanism allows the ACMPTC to react in real-time by reallocating resources dynamically to network congestion or high RTT that could affect AR/VR performance. In particular, the system adapts the congestion window \( cwnd_{i,j}[t] \) continuously based on real-time path conditions and behavior of each AR/VR stream \( S_j \), enabling efficient traffic management and smooth operation even when network conditions fluctuate

\subsubsection{Congestion Control}
For each AR/VR data stream \( S_j \) transmitted over \( P_i \), $cwnd_{i,j}[t]$, at $t$ governs how much data that AR/VR can transmit without receiving an ACK from the receiver. The congestion window is automatically tuned and adapted to network status (e.g., packet loss, RTT, congestion level, etc.) from the server side to prevent overcapacity of networks while guaranteeing an audiovisual message transmission in AR/VR application. The congestion window is then updated on each time step \( t \) as follows:
\begin{equation*}
  cwnd_{i,j}[t+1] = 
  \begin{cases} 
    cwnd_{i,j}[t] + \alpha \left(1 - \rho_i[t] - \sum_{j} \gamma_j \delta_j[t]\right), \\
    \quad \text{if } \rho_i[t] < \rho_{\text{th}}, \\
    \max \left( cwnd_{i,j}[t] - \beta  cwnd_{i,j}[t] \zeta, 1 \right), \\
    \quad \text{otherwise}.
  \end{cases}
\end{equation*}
where $\zeta= \rho_i[t] + \sum_{j} \gamma_j \delta_j[t] $ and \( \alpha \) is the scaling factor for the congestion window increase rate in the event of uncongestion (i.e., \( \rho_i[t] < \rho_{\text{th}} \)), \( \beta \) is the scaling factor that controls the rate of decrease of the congestion window when packet loss exceeds the threshold, \( \rho_i[t] \) is the packet loss rate on path \( P_i \) at time \( t \), \( \rho_{\text{th}} \) is the predefined threshold for acceptable packet loss on the path, \( \delta_j[t] \) is a metric representing deviation from typical traffic behavior for stream \( S_j \), accounting for unpredictable traffic surges or reductions specific to AR/VR interactions (e.g., sudden bursts of data when rendering complex scenes), and \( \gamma_j \) is a weighting factor that reflects the relative impact of the AR/VR stream's behavior on the congestion window.

\subsubsection{AR/VR-Specific Congestion Sensitivity}
Due to the interactive nature of AR/VR applications, the system places greater emphasis on latency and packet loss than traditional congestion control algorithms. For AR/VR streams, even small packet loss or excessive latency can disrupt the user experience by causing lag or dropped frames. To account for this, the congestion control mechanism adjusts the congestion window more aggressively for AR/VR data streams, especially when packet loss rate \( \rho_i[t] \) approaches the threshold \( \rho_{\text{th}} \) and \( RTT_i[t] \) increases significantly, indicating growing congestion on path \( P_i \). For AR/VR streams, the weighting factor \( \gamma_j \) is typically higher, making the system more sensitive to changes in network conditions, ensuring that the data flow can quickly adapt to maintain a high QoS for users.

\subsubsection{Congestion Window Adjustment}
AR/VR streams often experience fluctuations in data rate due to varying complexity in the content being streamed (e.g., rendering simple versus complex virtual scenes). This leads to changing bandwidth demands that must be handled dynamically. To accommodate these fluctuations, the system adjusts the congestion window \( cwnd_{i,j}[t] \) based on the real-time behavior of each AR/VR stream. At each time step \( t \), the system evaluates the current data rate and compares it to the expected data rate for the AR/VR stream. If the data rate exceeds the expected rate, the system increases the congestion window to accommodate the higher data demands. Conversely, if the data rate drops, the system reduces the window to avoid overloading the network. The adjustment is mathematically represented as:
\begin{equation}
  cwnd_{i,j}[t+1] = cwnd_{i,j}[t] + \delta \left( \frac{R_{S_j}[t]}{R_{{S_j}_e}[t]} - 1 \right),
\end{equation}
where \( \delta \) is a scaling factor that adjusts the congestion window proportionally to the difference between the actual \( R_{S_j}[t] \) and expected \( R_{{S_j}_e}[t] \) data rates.
ACMPTC system's congestion control mechanism maintains a high QoS for AR/VR users. The system prioritizes paths with the lowest latency and packet loss, even if those paths offer lower bandwidth. The feedback mechanism continuously monitors user satisfaction and network conditions to adjust the real-time congestion window, ensuring that the QoS remains high. If a particular path's congestion level \( C_i[t] \) exceeds a threshold \( C_{\text{th}} \), or if the user experience score \( U_{x_i}[t] \) drops, the system dynamically adjusts the path assignment and reduces the congestion window on that path to avoid further degradation in performance.

\subsection{Dynamic Bandwidth Allocation}
In the ACMPTC system, dynamic bandwidth allocation ensures that each AR/VR data stream \( S_j \) is provided with sufficient bandwidth based on its real-time requirements and the current network conditions. The system continuously monitors the available bandwidth on each path \( B_i[t] \) and adjusts the allocation to meet the performance demands of each stream.
At each time step \( t \), the bandwidth \( B_{i,j}[t] \) allocated to AR/VR stream \( S_j \) on path \( P_i \) is determined based on the available bandwidth \( B_i[t] \) and the congestion window \( cwnd_{i,j}[t] \). The dynamic allocation process is formulated as:
\begin{equation}
  B_{i,j}[t] = \min\left( \frac{cwnd_{i,j}[t]}{RTT_i[t]}, B_{i,\text{max}} - \sum_{j} u_{j,i}[t] \right),
\end{equation}
where \( cwnd_{i,j}[t] \) is the congestion window for stream \( S_j \) on path \( P_i \),
 \( RTT_i[t] \) is the round-trip time for path \( P_i \),
 \( B_{i,\text{max}} \) is the maximum bandwidth capacity of path \( P_i \),
 \( u_{j,i}[t] \) is the actual bandwidth used by stream \( S_j \) on path \( P_i \) at time \( t \).
The system seeks to allocate bandwidth based on the path's available capacity while ensuring that the AR/VR stream's real-time needs are met. The allocation is designed to adapt to  the congestion control mechanism and the network conditions, ensuring efficient bandwidth utilization across all streams.
When the path drops below a critical threshold, the system reallocates the traffic to other paths with more available capacity, ensuring that the AR/VR data streams receive sufficient bandwidth to maintain the QoS.
The system continuously monitors bandwidth utilization and dynamically reallocates traffic across multiple paths as required. If the bandwidth requirements of an AR/VR stream \( S_j \) exceed the available capacity of the assigned path(s), the system reallocates traffic to other paths \( P_j \) where bandwidth is available:
  \begin{equation}
  \mathcal{P}_{S_j}^*[t] = \arg\max_{\mathcal{P}_{S_j} \subseteq \mathcal{P}} \left\{ B_j[t] \right\},
\end{equation}
This dynamic reallocation ensures that no single path is overloaded while maintaining the overall performance for each AR/VR stream.

\subsection{Feedback Loop for Real-Time Optimization}
The feedback loop is a crucial component of the ACMPTC system, enabling real-time optimization based on current network conditions and QoS. It monitors each path's performance metrics, such as RTT, bandwidth, congestion level, and packet loss, and makes real-time adjustments to maintain optimal performance. The feedback metric \( F_{b_{i}}[t] \) at time \( t \) for path \( P_i \) is defined as:
\begin{equation}
  F_{b_{i}}[t] = \eta \cdot (\tau_i - U_{x_i}[t]),
\end{equation}
where \( \eta \) is the learning rate that controls the sensitivity of the system's adjustments, \( \tau_i \) is the desired performance level for path \( P_i \), and \( U_{x_i}[t] \) represents the user experience score for path \( P_i \) at time \( t \). 
This feedback mechanism dynamically adjusts network parameters—such as path selection, congestion window size \( cwnd_{i,j}[t] \), and bandwidth allocated to AR/VR stream \( S_j \) on path \( P_i \) \( B_{i,j}[t] \)—based on real-time network performance and user feedback. As a result, the system continuously optimizes the delivery of AR/VR streams, enhancing the QoS for users \cite{emptcp2024}.
AR/VR applications are highly susceptible to network performance, requiring high bandwidth, low RTT, low latency, and minimal packet loss to deliver a seamless experience. The ACMPTC system is designed to address these requirements by continuously adapting to the unique demands of AR/VR content. By utilizing multiple paths and dynamically adjusting RTT, congestion windows, and other resources, the system ensures that AR/VR streams maintain high levels of QoS \cite{emptcp2024}.

\subsection{Utility Function}
The ACMPTC system uses a utility function to quantify the performance of each AR/VR data stream \( S_j \) based on key network metrics such as bandwidth, latency, packet loss, and QoS. The utility function \( U_{S_j}[t] \) at time \( t \) for stream \( S_j \) is defined as:
\begin{equation}
  U_{S_j}[t] \!=\! w_B B_{S_j}[t] \!-\! w_L L_{S_j}[t] \!-\! w_P \rho_{S_j}[t] \!+\! w_Q  QoS_{S_j}[t],
\end{equation}
where \( B_{S_j}[t] \) is the effective bandwidth allocated to stream \( S_j \) at time \( t \),
\( L_{S_j}[t] \) is the latency experienced by stream \( S_j \) at time \( t \),
 \( \rho_{S_j}[t] \) is the packet loss rate for stream \( S_j \),
 \(  QoS_{S_j}[t] \) is the  QoS score for stream \( S_j \),
 \( w_B, w_L, w_P, w_Q \) are weighting factors prioritizing the importance of bandwidth, latency, packet loss, and  QoS, respectively.
The goal of the utility function is to maximize the performance of the AR/VR stream by optimizing the allocation of network resources. The system continuously adjusts network parameters to maximize \( U_{S_j}[t] \), ensuring that the stream experiences high bandwidth, low latency, minimal packet loss, and an optimal  QoS.

\subsection{Problem Formulation}
The problem formulation for the ACMPTC system aims to optimize the performance of AR/VR data streams by addressing key network metrics such as bandwidth, latency, packet loss, and QoS. 
The optimization problem is mathematically expressed as:
\begin{subequations}\label{Eq_op_2}
\begin{align}
    & \mathop{\max}\limits_{\mathcal{P}_{S_j}[t], B_{i,j}[t], cwnd_{i,j}[t]} \sum_{j=1}^{n} \sum_{t=0}^{T} \Big( w_B B_{S_j}[t] - w_L L_{S_j}[t] \nonumber\\
    & \quad\quad\quad\quad\quad\quad\quad - w_P \rho_{S_j}[t] + w_Q  QoS_{S_j}[t] \Big) \label{Eq_ob_2} \\
    & \text{s.t.}\ \sum_{j=1}^{k} B_{i,j}[t] \leq B_{i,\text{max}}, \quad \forall i \label{Eq_op2_c1}\\
    & \!cwnd_{i,j}[t\!+\!1]\! \!= \!\!
      \begin{cases} 
    \!\!cwnd_{i,j}[t] \!+\! \alpha\!\! \left(\!1 \!-\! \rho_i[t] \!-\! \!\sum_{j} \! \gamma_j\!  \delta_j[t]\right)\!  \\
    \quad \text{if } \rho_i[t] < \rho_{\text{th}}, \\
    \!\! \max \left( cwnd_{i,j}[t] \!-\! \beta  cwnd_{i,j}[t] \zeta, 1 \right) \\
    \quad \text{otherwise}.
  \end{cases} \!\!\label{Eq_op2_c2} \\
    & L_{S_j}[t] \leq L_{\text{max}}, \quad \forall j \label{Eq_op2_c3} \\
    & \rho_{S_j}[t] \leq \rho_{\text{max}}, \quad \forall j \label{Eq_op2_c4} \\
    &  QoS_{S_j}[t] \geq  QoS_{\text{min}}, \quad \forall j \label{Eq_op2_c5} \\
    & \mathcal{P}_{S_j}^*[t]\! =\! \arg\max_{\mathcal{P}_{S_j}\! \subseteq \mathcal{P}} \Bigg( \!\!\sum_{P_i \in \mathcal{P}_{S_j}} \!\!\! \Big(\! \alpha B_i[t] \!-\! \beta L_i[t] \!-\! \gamma \rho_i[t] \Big)\!\! \Bigg) \label{Eq_op2_c6}
\end{align}
\end{subequations}

The objective function in (\ref{Eq_ob_2}) seeks to maximize the utility \( U_{S_j}[t] \) for each AR/VR stream over time \( T \), ensuring optimal performance. The path assignment constraint in (\ref{Eq_op2_c6}) ensures that each stream \( S_j \) is assigned to an optimal subset of paths \( \mathcal{P}_{S_j}[t] \subseteq \mathcal{P} \), selected based on real-time network conditions. The bandwidth constraint in (\ref{Eq_op2_c1}) ensures that the allocated bandwidth \( B_{i,j}[t] \) does not exceed the maximum available bandwidth \( B_{i,\text{max}} \) for each path \( P_i \). The congestion control constraint in (\ref{Eq_op2_c2}) dynamically adjusts the congestion window based on network conditions. The latency and packet loss constraints in (\ref{Eq_op2_c3}) and (\ref{Eq_op2_c4}) ensure acceptable performance levels, while the QoS constraint in (\ref{Eq_op2_c5}) guarantees a minimum level of user satisfaction.

\section{DRL-based Multi-Agent Solution for ACMPTC} \label{Section:ProposedSolution}

The optimization problem formulated in (\ref{Eq_op_2}) for the ACMPTC model aims to maximize the QoS for AR/VR data streams by balancing key network metrics such as bandwidth, latency, and packet loss. To solve this problem in a distributed and scalable manner, we adopt a \textit{Multi-Agent Advantage Actor-Critic (A2C)} algorithm. Each agent manages a specific AR/VR stream, learning optimal policies for dynamic decision-making in a decentralized manner. This multi-agent framework allows agents to operate independently but collaboratively to optimize overall network performance.

\subsection{Multi-Agent Architecture Overview}
Each AR/VR stream \( S_j \) is assigned to an individual DRL agent responsible for dynamically managing path selection, congestion control, and bandwidth allocation. The agents share a familiar environment (the network) but make decisions independently based on their localized observations.

\textit{Decentralized Path Assignment and Reallocation:} Each agent determines the optimal paths \( \mathcal{P}_{S_j}^*[t] \) based on real-time local observations of congestion, RTT, and bandwidth availability. Each agent makes Path assignment decisions independently, allowing for better scalability.
    
 \textit{Independent Bandwidth Management:} Each agent dynamically adjusts \( B_{i,j}[t] \) for the stream it manages, ensuring efficient bandwidth allocation based on current requirements and network conditions observed by the agent.
    
 \textit{Adaptive Congestion Control:} Agents manage the congestion window \( cwnd_{i,j}[t] \) for their assigned streams independently, ensuring that packet loss is minimized while maintaining the required QoS.
This multi-agent approach ensures that each AR/VR stream is managed in an adaptive, decentralized manner, optimizing performance without relying on a centralized agent, thus improving scalability and robustness.

\subsection{State Representation}
Each agent observes its local state independently, encapsulating relevant network conditions and user experience metrics. The state vector for agent \( j \), \( s_{j,t} \), at time \( t \), is defined as:
\begin{equation}
s_{j,t} = \left( B_i[t], L_i[t], \rho_i[t], F_{j,t}, U_{x_{j,t}}, C_i[t] \right) \quad \forall i \in \mathcal{P}
\end{equation}
where \( B_i[t] \) is the available bandwidth on path \( P_i \), \( L_i[t] \) represents the latency on path \( P_i \), \( \rho_i[t] \) is the packet loss rate, \( F_{j,t} \) is the feedback metric for stream \( S_j \), \( U_{x_{j,t}} \) is the user experience score, and \( C_i[t] \) captures the congestion level on path \( P_i \). 

\subsection{Action Space}
The action space for each agent \( j \), \( a_{j,t} \), at time \( t \), includes:

\textit{Path Selection:} Choosing the optimal subset of paths \( \mathcal{P}_{S_j}^*[t] \) for each AR/VR stream \( S_j \).

\textit{Bandwidth Allocation:} Adjusting \( B_{i,j}[t] \) based on the current network capacity and bandwidth requirements of the streams.

\textit{Congestion Window Adjustment:} Modifying \( cwnd_{i,j}[t] \) to control data flow rates and minimize packet loss.

\subsection{Reward Function}
The reward function \( R(s_{j,t}, a_{j,t}) \) provides feedback to each agent, guiding them toward decisions that optimize QoS. The reward for agent \( j \) at time \( t \) is:
\begin{equation}
R(s_{j,t}, a_{j,t}) = w_B B_{S_j}[t] - w_L L_{S_j}[t] - w_P \rho_{S_j}[t] + w_Q QoS_{S_j}[t]
\end{equation}
where \( w_B, w_L, w_P, \) and \( w_Q \) are weighting factors prioritizing bandwidth, latency, packet loss, and QoS, respectively.

\subsection{Actor-Critic Networks for Each Agent}
Each agent employs its own actor and critic networks. The actor network determines the policy \( \pi_j(a_{j,t} | s_{j,t}; \theta^{\pi}_j) \), while the critic network estimates the value function \( V_j(s_{j,t}; \theta^{V}_j) \).

\textit{Actor Network:}
    \begin{equation}
    \pi_j(a_{j,t} | s_{j,t}; \theta^{\pi}_j) = \text{softmax}\left( \text{NN}_{\text{actor}}(s_{j,t}; \theta^{\pi}_j) \right)
    \end{equation}

\textit{Critic Network:}
    \begin{equation}
    V_j(s_{j,t}; \theta^{V}_j) = \text{NN}_{\text{critic}}(s_{j,t}; \theta^{V}_j)
    \end{equation}

\subsection{Objective Function}
The objective function for each agent is to maximize the expected return by learning the policy and value functions:
\begin{equation}
L(\theta^{\pi}_j, \theta^{V}_j) = \mathbb{E}_{\pi_j}\left[ \sum_{t=0}^{T} \gamma^t A_{j,t}(s_{j,t}, a_{j,t}) \right]
\end{equation}
where the advantage function \( A_{j,t}(s_{j,t}, a_{j,t}) \) is defined as:
\begin{align}
A_{j,t}(s_{j,t}, a_{j,t}) &= R(s_{j,t}, a_{j,t}, s_{j,t+1}) + \gamma V_j(s_{j,t+1}; \theta^{V}_j) \nonumber \\
&\quad - V_j(s_{j,t}; \theta^{V}_j)
\end{align}

The advantage function \( A_t \) measures the relative value of taking action \( a_t \) in state \( s_t \), guiding the agent toward actions that contribute to maximizing the expected return, thereby improving QoS for AR/VR streams.

\subsection{Training Process}
During the training phase, the actor and critic network parameters, \( \theta^{\pi} \) and \( \theta^{V} \), are iteratively updated to maximize the expected return. The optimization is performed by taking the gradients of the objective function concerning \( \theta^{\pi} \) and \( \theta^{V} \):
\begin{align}
\nabla_{\theta^{\pi}} L(\theta^{\pi}) &= \mathbb{E}_{\pi}\left[ \nabla_{\theta^{\pi}} \log \pi(a_t \mid s_t; \theta^{\pi}) A_t(s_t, a_t) \right] \\
\nabla_{\theta^{V}} L(\theta^{V}) &= \mathbb{E}_{\pi}\left[ \left(R + \gamma V(s_{t+1}; \theta^{V}) - V(s_t; \theta^{V})\right)^2 \right]
\end{align}
The actor-network is updated by maximizing the expected advantage, which helps the policy learn to select actions that improve network performance. The critic network is updated by minimizing the temporal difference error, improving the accuracy of the value estimation for future rewards.

\subsection{ A2C Algorithm}
The A2C algorithm for the ACMPTC model iteratively updates the policy and value functions to enhance decision-making capabilities under dynamic AR/VR multimedia application demands. Algorithm~\ref{Algorithm} outlines the training process:
\begin{algorithm}
\caption{Multi-Agent A2C Algorithm for Optimizing ACMPTC Model}
\label{Algorithm}
\begin{algorithmic}[1]
\For{each agent \( j = 1, \ldots, n \)}
  \State Initialize actor network parameters \( \theta^{\pi}_j \) and critic network parameters \( \theta^{V}_j \)
  \State Initialize the environment state \( s_{j,0} \)
\EndFor
\For{each episode \( e = 1, \ldots, E \)}
  \For{each agent \( j = 1, \ldots, n \)}
    \For{each time step \( t = 0, \ldots, T \)}
      \State Observe the current state \( s_{j,t} \)
      \State Select action \( a_{j,t} \) from \( \pi_j(a_{j,t} | s_{j,t}; \theta^{\pi}_j) \)
      \State Execute action \( a_{j,t} \) and observe reward \( r_{j,t} \) and new state \( s_{j,t+1} \)
      \State Calculate the advantage: \( A_{j,t} = r_{j,t} + \gamma V_j(s_{j,t+1}; \theta^{V}_j) - V_j(s_{j,t}; \theta^{V}_j) \)
      \State Update critic by minimizing the loss: \( \mathcal{L}^{\text{critic}}_j = (A_{j,t})^2 \)
      \State Update actor by maximizing the expected reward using \( \nabla_{\theta^{\pi}_j} \log \pi_j(a_{j,t} | s_{j,t}; \theta^{\pi}_j) A_{j,t} \)
      \State Apply gradient descent to adjust parameters \( \theta^{\pi}_j \) and \( \theta^{V}_j \)
      \State Update \( s_{j,t} \gets s_{j,t+1} \)
    \EndFor
  \EndFor
\EndFor
\end{algorithmic}
\end{algorithm}

\subsection{Computational Complexity}
The computational complexity of the multi-agent A2C algorithm primarily depends on the size of the state and action spaces, as well as the complexity of the neural network operations for each agent. Assuming fully connected neural networks, the complexity can be expressed as:
\begin{equation}
\mathcal{O}(n \cdot I \cdot (c_1 \cdot |S| \cdot N + c_2 \cdot |A| \cdot N))
\end{equation}
where \( n \) is the number of agents, \( |S| \) is the cardinality of the state space, \( |A| \) is the cardinality of the action space, \( N \) is the number of parameters in the actor and critic networks, \( I \) is the number of iterations, and \( c_1 \) and \( c_2 \) are constants representing the computational cost of forward and backward passes in the networks.

\section{Simulation and Evaluation} \label{SE}
In this section, we provide the results of our simulation experiments for ACMPTC model based on 360-degree video streaming dataset \cite{corbillon2017360}. Explicitly focusing on AR/VR streaming environments, we evaluate the performance of ACMPTC concerning throughput, latency, and packet loss, along with overall QoS parameters influencing end-to-end media experience. The performance analysis is done using different topologies and also for non-uniform traffic effects.
We employed a 360-degree video dataset to evaluate the ACMPTC approach and artificially created network scenarios similar to those encountered by streaming environments of AR/VR experiences. We selected the simulation parameters to represent real-world variability for such applications' bandwidth, latency, and packet loss. The average throughput, packet loss, and latency were measured in each experiment under different network conditions. The optimum results are compared with the standard MPTCP, TCP, and ACMTPC under different network conditions.

\subsection{Implementation of DRL}
Details on the technical setup for implementing DRL, including training processes, computational resources, and data management, ensuring scalability and responsiveness to AR/VR needs.
The efficacy of the ACMPTC framework is rigorously tested through a comprehensive simulation to evaluate its performance in dynamically adapting to fluctuating bandwidth requirements typical of AR/VR streaming applications. 
The Actor-Critic architecture facilitates efficient policy learning, with the actor proposing actions and the critic evaluating their potential based on the current policy and receiving the reward.
This section describes the simulation framework and parameters devised to simulate a realistic network environment for evaluating the ACMPTC system, focusing on optimizing AR/VR streaming network paths. It includes analyzing traffic pattern variability using the dataset, with parameters detailed in Table~\ref{tab:simulation_parameters}. 
\begin{table}[ht]
\centering
\caption{Simulation Setup for ACMPTC System}
\begin{tabular}{l|l}
\hline
\textbf{Parameter} & \textbf{Value} \\
\hline
Number of Paths ($n$) & 5 \\
Path Bandwidth Range ($B_{\text{min}}, B_{\text{max}}$) & 10 Mbps to 100 Mbps \\
Path Latency Range ($L_{\text{min}}, L_{\text{max}}$) & 10 ms to 100 ms \\
Packet Loss Rate & 0\% to 5\% \\
Learning Rate ($\alpha$) & 0.01 \\
Discount Factor ($\gamma$) & 0.95 \\
Exploration Rate ($\epsilon$) & Starts at 1.0, decaying to 0.01 \\
Reward Weights: Throughput ($\alpha$) & 0.7 \\
Reward Weights: Latency ($\beta$) &0.2 \\
Reward Weights: Packet Loss ($\gamma$) & 0.1 \\
Total Simulation Time & 1000 seconds \\
Time Step ($\Delta_i t$) & 1 second \\
AR/VR Streaming Data Rate & 5 Mbps to 50 Mbps \\
Background Traffic Rate & 0 Mbps to 50 Mbps \\
Signal Segment Length for FFT & 1024 samples \\
\hline
\end{tabular}
\label{tab:simulation_parameters}
\end{table}

\subsection{Dataset: 360-Degree Video Streaming Dataset}
For simulating the ACMPTC model, we used the 360-degree Video Streaming dataset \cite{corbillon2017360}. Moreover, it simulates realistic video streaming conditions for testing high-bandwidth, low-latency AR/VR applications [3]. This dataset includes 360-degree VR video traces steamed in all network conditions, so it is highly suitable for evaluating various dynamic networking ACMPTC protocols.
The dataset contains features necessary to simulate the ACMPTC model.
It contains multiple video bitrates and resolutions: from 1 Mbps to 20 Mbit/s at scale, ranging on different types of videos (resolution goes from standard-definition up until ultra-high-def), which is crucial to exercise high-bandwidth AR/VR scenarios where our ACMPTC model has been enabling bandwidth scales.
The traces span between 5 and 100Mbps bandwidth fluctuations and latency values ranging from approximately tens to hundreds of milliseconds — critical components for the model that allow it to show dynamic path selection optimized for low-latency-sensitive applications (e.g., AR/VR).
The packet loss data varies from 0\%-5\% with varying network conditions so that the ACMPTC model can test how it responds to real-world commonly occurring disruptions and design efficient mechanisms for handling different levels of packet losses.
It contains the viewport traces of the user (head movement and focused parts on 360 video during playback) as a dataset. This characteristic facilitates the model in dynamically changing bandwidth allocation according to user activity and focusing on delivering higher resolution in focus regions: a must-have for AR/VR experiences that contribute so much towards better QoE.
Time-series data is helpful because it models the temporal variability of network conditions and user interactions, which are also necessary for testing how well ACMPTC can adapt in real time.
These features enable a thorough testing of the ACMPTC model concerning its bandwidth allocation dynamic adaptability, path selection optimality, and congestion control. The large number of video bitrates, network bandwidth swings, and latency in the dataset make it an ideal platform to evaluate high-load, low-latency streaming control by the model. In addition, by providing packet loss data and user viewport traces to the model that optimizes network resource allocation concerning real-time user interactions, better QoS/QoE should be achieved for AR/VR streaming.

\subsection{Quantitative Results Analysis}
\begin{figure*}
  \centering
  \begin{subfigure}[H]{0.20\textwidth}
    \centering
  \includegraphics[width=\textwidth]{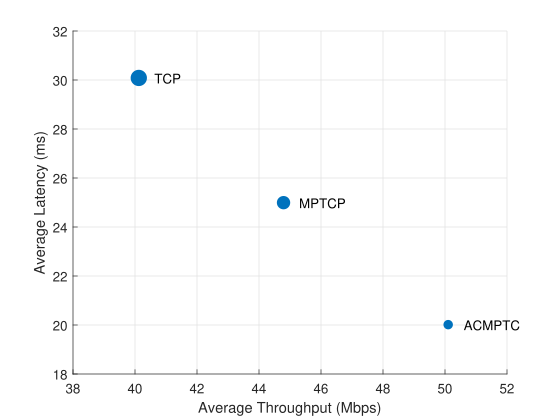}
  \caption{Packet loss visualization}
  \label{fig:bubble_packet_loss}
  \end{subfigure}
  \hfill
  \begin{subfigure}[H]{0.20\textwidth}
    \centering
  \includegraphics[width=\textwidth]{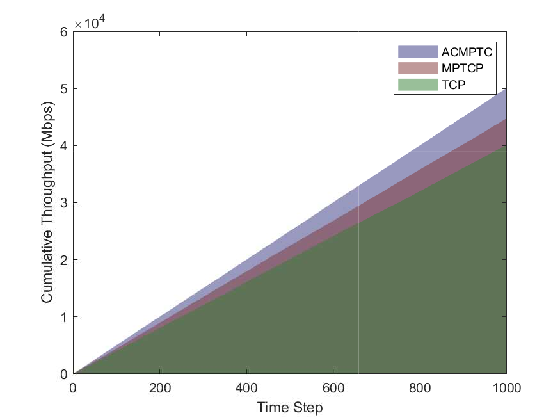}
  \caption{Throughput over time}
  \label{fig:cumulative_throughput}
  \end{subfigure}
  \hfill
  \begin{subfigure}[H]{0.20\textwidth}
    \centering
  \includegraphics[width=\textwidth]{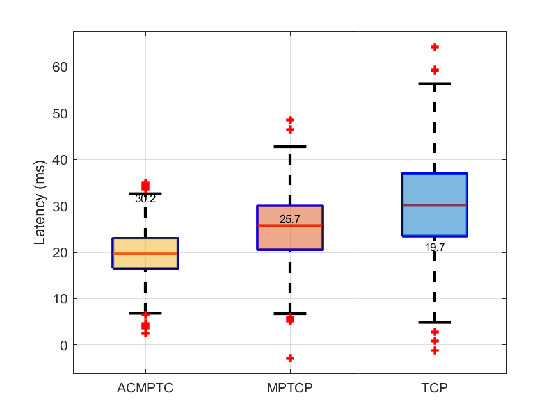}
  \caption{Latency protocol}
  \label{fig:latency_protocol_comparison}
  \end{subfigure}
  \hfill
  \begin{subfigure}[H]{0.20\textwidth}
    \centering
    \includegraphics[width=\textwidth]{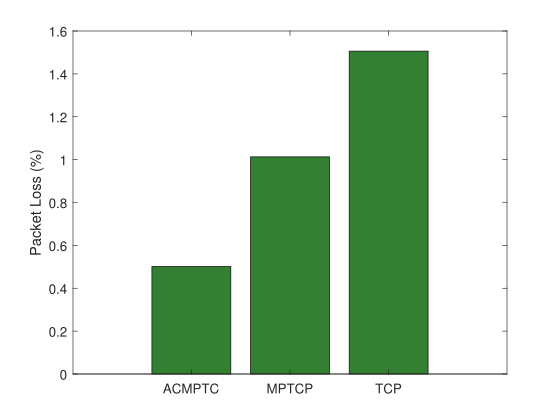}
    \caption{Packet loss rates}
    \label{fig:packet_loss_rates}
  \end{subfigure}
\caption{Network metrics analysis}
\label{fig:network_metrics_analysis}
\end{figure*}

\begin{figure*}
  \centering
  \begin{subfigure}[H]{0.20\textwidth}
    \centering
    \includegraphics[width=\textwidth]{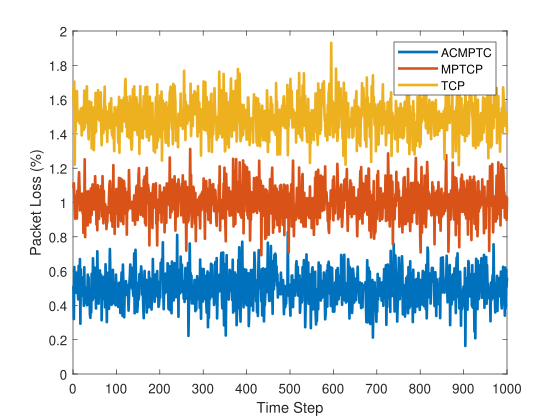}
    \caption{Packet loss over time}
    \label{fig:packet_loss_over_time}
  \end{subfigure}
  \hfill
  \begin{subfigure}[H]{0.20\textwidth}
    \centering
    \includegraphics[width=\textwidth]{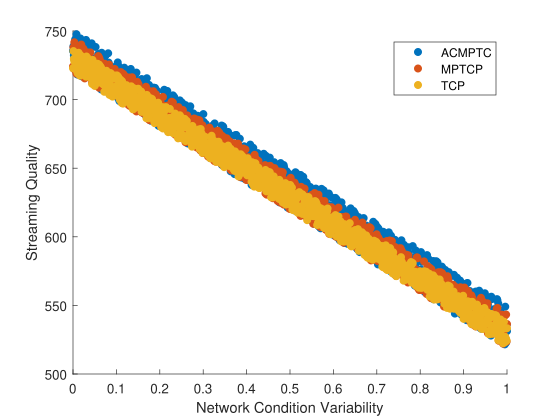}
    \caption{Streaming quality}
    \label{fig:streaming_quality_analysis}
  \end{subfigure}
  \hfill
  \begin{subfigure}[H]{0.20\textwidth}
    \centering
    \includegraphics[width=\textwidth]{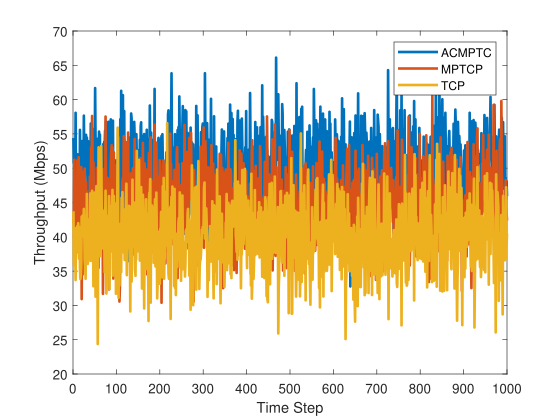}
    \caption{Throughput comparison}
    \label{fig:throughput_comparison}
  \end{subfigure}
  \hfill
  \begin{subfigure}[H]{0.20\textwidth}
    \centering
    \includegraphics[width=\textwidth]{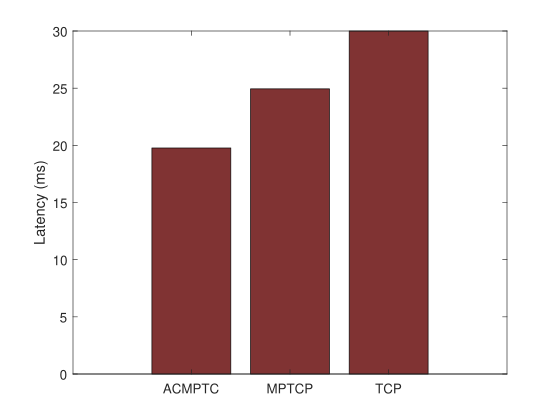}
    \caption{Latency across protocols}
    \label{fig:latency_protocols}
  \end{subfigure}
\caption{Network performance insights}
\label{fig:network_performance_insights}
\end{figure*}

\begin{figure*}
  \centering
  \begin{subfigure}[H]{0.20\textwidth}
    \centering
    \includegraphics[width=\textwidth]{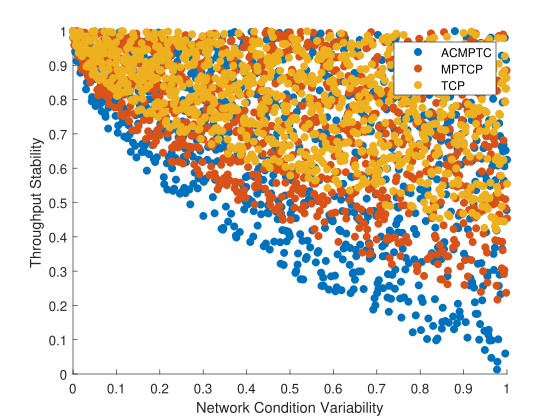}
    \caption{Throughput stability}
    \label{fig:throughput_stability_variable_conditions}
  \end{subfigure}
  \hfill
  \begin{subfigure}[H]{0.20\textwidth}
    \centering
    \includegraphics[width=\textwidth]{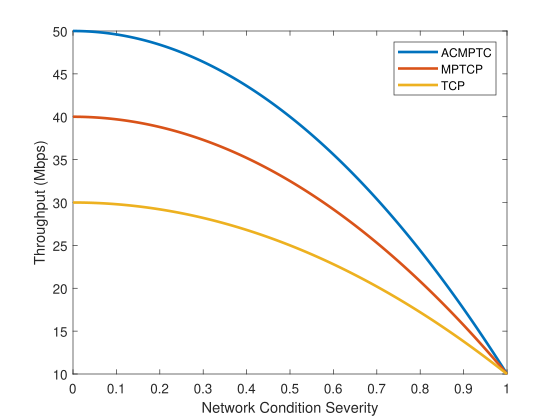}
    \caption{At extreme conditions}
    \label{fig:throughput_extreme_conditions}
  \end{subfigure}
  \hfill
  \begin{subfigure}[H]{0.20\textwidth}
    \centering
    \includegraphics[width=\textwidth]{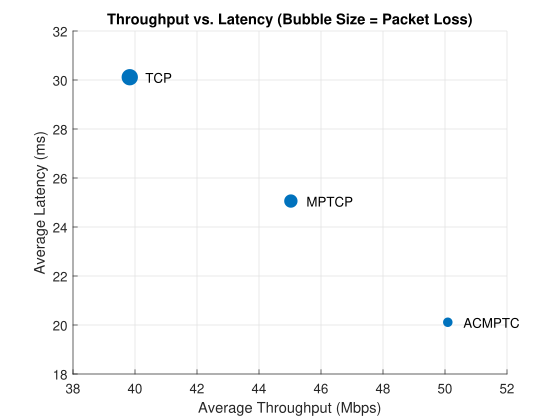}
    \caption{Throughput \& latency}
    \label{fig:throughput_vs_latency}
  \end{subfigure}
  \hfill
  \begin{subfigure}[H]{0.20\textwidth}
    \centering
    \includegraphics[width=\textwidth]{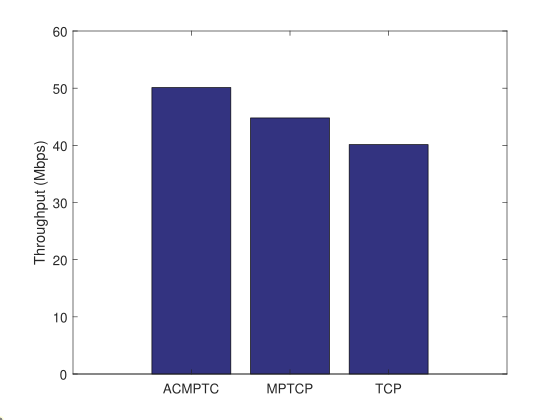}
    \caption{Throughput efficiency}
    \label{fig:throughput_efficiency}
  \end{subfigure}
\caption{Network dynamics examination}
\label{fig:network_dynamics_examination}
\end{figure*}

Our comprehensive analysis evaluates various network protocols to understand their performance characteristics under different conditions. Central to our investigation is comparing critical performance metrics such as latency, packet loss, and throughput.
Using a bubble-size visualization technique, we visually examine packet loss across the network in Fig.~\ref{fig:bubble_packet_loss}. 
Further analysis of the cumulative throughput achieved over time is depicted in Fig.~\ref{fig:cumulative_throughput}. This graph emphasizes the protocols' capability to manage data transmission efficiently, showcasing their resilience and robustness.
The investigation into latency distributions, shown in Fig.~\ref{fig:latency_protocol_comparison}, reveals how different protocols manage latency. The variability across protocols sheds light on the trade-offs between speed and reliability in network communications.
Our study also tracks the evolution of packet loss over time, as illustrated in Fig.~\ref{fig:packet_loss_over_time}. This analysis offers insights into the temporal stability of the protocols, highlighting their performance consistency.
Fig.~\ref{fig:streaming_quality_analysis} focuses on streaming quality, a crucial aspect of QoS in networked applications that evaluate the impact of protocol selection on streaming services' quality, underlining the importance of protocol efficiency for end-user satisfaction.
In Fig.~\ref{fig:throughput_comparison}, we conduct a comparative throughput analysis over time, further delineating the throughput capabilities of each protocol. 
To assess the robustness of network protocols under diverse conditions, we examine throughput stability in variable network environments (Fig.~\ref{fig:throughput_stability_variable_conditions}) and under extreme network conditions (Fig.~\ref{fig:throughput_extreme_conditions}). These figures collectively underscore the adaptability of the protocols, demonstrating their suitability across a wide range of operational scenarios.
Lastly, Fig.~\ref{fig:throughput_vs_latency} explores the relationship between throughput and latency, providing insights into the complex dynamics that govern the balance between speed and reliability in network operations.
The analyses from Fig.~\ref{fig:bubble_packet_loss} to Fig.~\ref{fig:throughput_vs_latency} contribute to a nuanced understanding of network protocol performance. 
\subsection{Discussion on AR/VR Streaming Performance}
Our analysis, centered around Fig.~\ref{fig:bubble_packet_loss} through Fig.~\ref{fig:throughput_efficiency}, sheds light on the performance of ACMPTC, MPTCP, and TCP protocols in the context of AR/VR streaming. These insights are critical for developers and network engineers aiming to optimize AR/VR experiences.
Fig.~\ref{fig:streaming_quality_analysis} addresses streaming quality, a paramount concern in AR/VR multimedia applications. The graph reveals how each protocol's handling of data transmission impacts the overall quality of streaming content. Notably, protocols that maintain higher throughput and lower latency, as shown in Fig.~\ref{fig:cumulative_throughput} and Fig.~\ref{fig:latency_protocol_comparison}, are better suited for AR/VR streaming, where even minor disruptions can significantly detract from the QoS.
The stability of throughput under variable network conditions, explored in Fig.~\ref{fig:throughput_stability_variable_conditions}, becomes especially relevant in AR/VR multimedia applications. These applications demand high performance under ideal conditions and robustness against network quality fluctuations. Similarly, the protocols' behavior under extreme network conditions, depicted in Fig.~\ref{fig:throughput_extreme_conditions}, underscores the importance of resilience in maintaining an immersive AR/VR experience.
Moreover, the trade-off between throughput and latency, analyzed in Fig.~\ref{fig:throughput_vs_latency}, highlights a crucial AR/VR streaming consideration. 

\section{Conclusion} \label{C}
This study examines the performance of ACMPTC, MPTCP, and TCP protocols against the backdrop of AR and VR technologies' growing demands, which require robust, agile networks capable of high bandwidth and low latency for optimal experiences. Our analysis identifies these protocols' varied performance and trade-offs under different conditions, emphasizing the necessity for efficiency, resilience, and adaptability in network protocol selection, especially for seamless AR/VR streaming. We highlight the critical ability of these protocols to adapt to fluctuating network conditions dynamically, ensuring consistent performance crucial for AR/VR multimedia applications. The paper contributes valuable insights into network protocol performance, establishing a foundation for future research to develop adaptive protocol mechanisms, explore efficient network architectures, and employ machine learning for network condition prediction and real-time optimization to advance network management for next-generation digital applications.

\bibliographystyle{IEEEtran}
\bibliography{IEEEabrv, Reference/mybib}

\end{document}